\pdfoutput=1
\documentclass[12pt]{article}
\usepackage{graphicx}
\usepackage{amssymb}
\usepackage{hyperref}
\usepackage[numbers,comma,square,sort&compress]{natbib}

\hypersetup{%
    pdftitle = {On the energy deposition by electrons in air and the accurate determination of the air-fluorescence yield},
    pdfsubject = {preprint},
    pdfkeywords = {air-fluorescence yield, energy deposition, Monte Carlo, Geant4, fluorescence telescopes, ultra-high-energy cosmic rays},
    pdfauthor = {J. Rosado, P. Gallego, D. García-Pinto, F. Blanco and F. Arqueros},
    pdfstartview = {FitH},
    }
\begin{document}
\begin{titlepage}
\begin{center}
\Huge {\bf On the energy deposition by electrons in air and the accurate determination of the air-fluorescence yield}
\par
\vspace*{2.0cm} \normalsize {\bf {J.~Rosado, P.~Gallego, D.~Garc\'{i}a-Pinto, F.~Blanco and F.~Arqueros}}
\par
\vspace*{0.5cm} \small \emph{Departamento de F\'{i}sica At\'{o}mica, Molecular y Nuclear, Facultad de Ciencias
F\'{i}sicas, Universidad Complutense de Madrid, E-28040 Madrid, Spain}
\end{center}
\vspace*{2.0cm}

\begin{abstract}
	The uncertainty in the absolute value of the air-fluorescence yield still puts a severe limit on the accuracy in
the primary energy of ultra-high-energy cosmic rays. The precise measurement of this parameter in laboratory is in turn
conditioned by a careful evaluation of the energy deposited in the experimental collision chamber. In this work we
discuss on the calculation of the energy deposition and its accuracy. Results from an upgraded Monte Carlo algorithm
that we have developed are compared with those obtained using Geant4, showing excellent agreement. These updated
calculations of energy deposition are used to apply some corrections to the available measurements of the absolute
fluorescence yield, allowing us to obtain a reliable world average of this important parameter.

\end{abstract}
\end{titlepage}

\section{Introduction}
\label{intro}

The fluorescence technique for detection of extensive air showers has been proved to be very fruitful. It is based on
the measurement of the faint fluorescence radiation of molecular nitrogen in the $\sim~300 - 400$~nm spectral range
induced by charged particles (mostly electrons) of a shower. By means of imaging telescopes, the longitudinal profile
of the shower is recorded allowing its geometrical reconstruction. In addition, it provides a nearly calorimetric
measure of the energy of the primary particle by means of the air-fluorescence yield $Y$, that is, the number of
photons per unit deposited energy. This key parameter is a function of wavelength and depends on the atmospheric
parameters, i.e., pressure, temperature and air composition. Nonetheless the energy scale of fluorescence telescopes is
mainly determined by its absolute value, usually given at standard atmospheric conditions (e.g., dry air at 1013~hPa
and 293~K) and for the most intense band of the spectrum, located at 337~nm.

The air-fluorescence yield is measured in dedicated experiments in laboratory using electron beams to excite air at
known conditions in a collision chamber. The induced fluorescence is registered by an appropriate optical system.
Several experiments use electrons from a $^{90}$Sr-$^{90}$Y radioactive source (average energy around 1~MeV), whereas
other experiments use beams from accelerators or electron guns (energies ranging from keV to GeV). It has been shown
that the fluorescence yield is nearly independent of electron energy~\cite{NJP,FLASH_thick,AIRFLY_E}, which is the
underlying principle of a calorimetric determination of the primary energy of extensive air showers. In fact, it is
also expected to be independent of the type of incident particle, since both fluorescence emission and energy
deposition are governed by low-energy secondary electrons produced in ionization processes.

An accurate experimental determination of the absolute air-fluorescence yield requires an end-to-end calibration of the
optical system as well as a careful Monte Carlo (MC) evaluation of the energy deposited in the field of view. We
discuss on the calculation of the energy deposition and its impact on the determination of the absolute
air-fluorescence yield in dedicated experiments. Results of energy deposition from an upgraded MC algorithm that we
have developed are compared with those obtained using Geant4. Finally, corrections previously proposed
in~\cite{AstroPh,UHECR11,ICRC11} to some fluorescence-yield measurements have been updated according to these
calculations, and a reliable world average of the absolute air-fluorescence yield is presented.

\section{Calculation of energy deposition}
\label{sec:energy}

\begin{figure}
\centering
\includegraphics[width=0.65\linewidth]{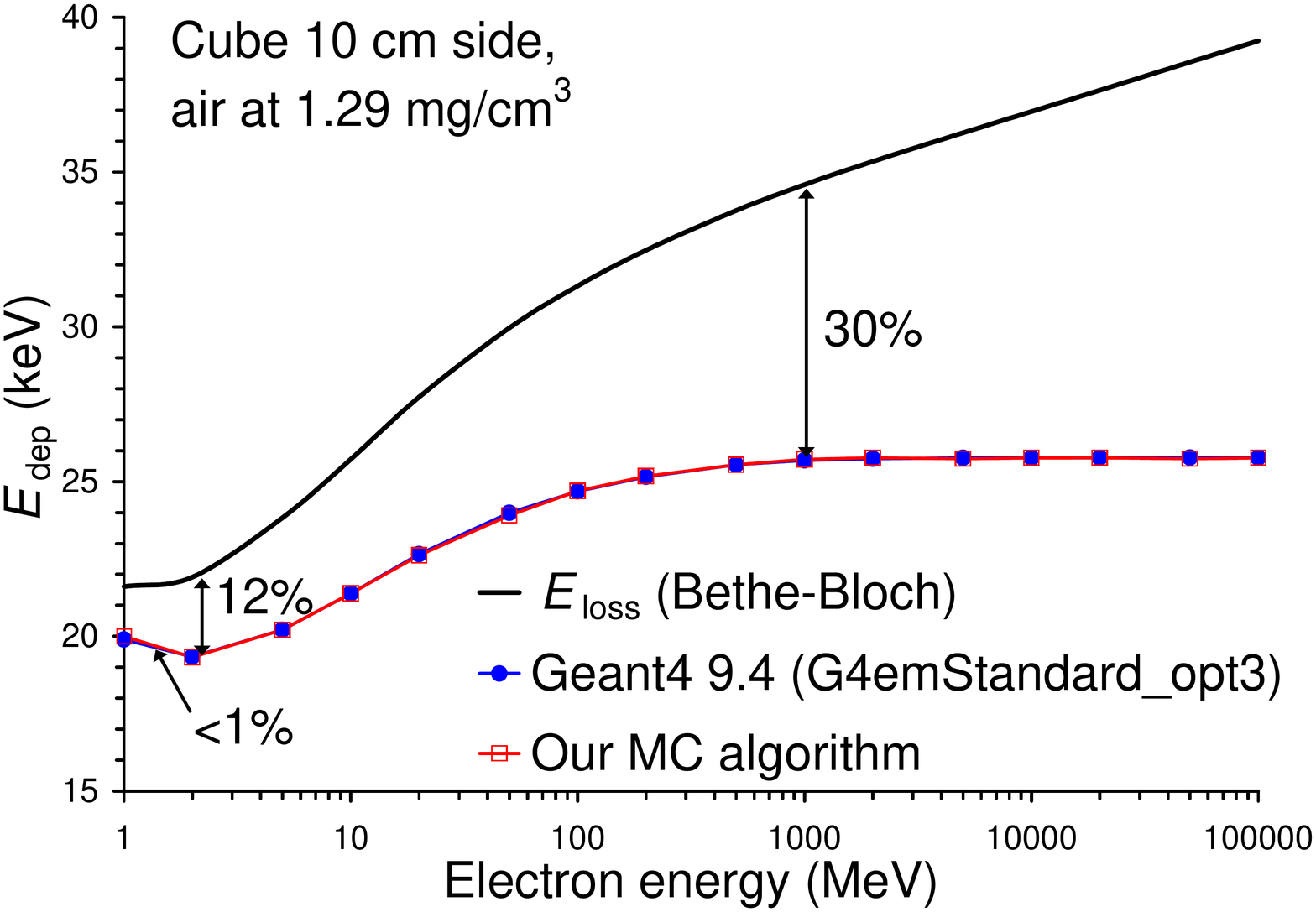}
\caption{Comparison of results of electron energy deposition from our upgraded MC simulation and Geant4 for a cube
10~cm side filled with air at 1.29~mg/cm$^3$.
An excellent agreement ($<1\%$) is observed in the wide energy range from 1~MeV to 100~GeV.
The energy loss given by the Bethe-Bloch formula is also shown to illustrate the contribution of escaping $\delta$
rays.}
\label{fig:cube}
\end{figure}

Collision stopping power is accurately described by the Bethe-Bloch theory, reference data being tabulated
in~\cite{ICRU37} for a number of materials including air. However, for a correct determination of the energy deposited
in a finite air volume, the fraction of energy transferred to secondary radiation escaping this volume has to be
evaluated. In particular, high-energy secondary electrons ($\delta$ rays) have large ranges. For instance, 100~keV
electrons can travel 14~cm in air at atmospheric pressure before being stopped. Although $\delta$ rays are barely
produced, they carry a significant fraction of the energy lost by the primary particle. This effect causes the energy
deposition to be lower than the total energy loss by $\sim8\%$ for incident electrons of 1~MeV and by more than $30\%$
for GeV electrons at usual experimental conditions (i.e., atmospheric pressure and a cm-sized collision chamber). This
is illustrated in figure~\ref{fig:cube}, where MC calculations of the energy deposition $E_{\rm dep}$ in a 10~cm cube
as a function of electron energy are plotted together with the Bethe-Bloch energy loss.

In addition, other features related to the propagation of electrons through the experimental collision chamber have
minor effects on the actual energy deposition. Firstly, stopping power varies along the trajectories of electrons as
they lose energy gradually. In the second place, scattering due to both elastic and inelastic collisions affects the
average path length of electrons inside the considered volume, contributing to either a rise or a reduction of the
energy deposition depending on geometry.

The energy deposition in the collision chamber of fluorescence-yield experiments is usually determined by available MC
codes like Geant4~\cite{Geant4} and EGS4~\cite{EGS4}. We have developed an alternative MC algorithm~\cite{NJP,Thesis}
using our own compilation of cross sections and other molecular data. Unlike standard MC codes, which normally employ
the multiple scattering approximation, our program simulates all the individual interactions of electrons. Comparison
of these independent simulations is thus a valuable test for the correctness and consistency of results on energy
deposition, allowing estimation of uncertainties. This is mandatory to achieve high precision level ($\lesssim5\%$) in
the absolute fluorescence yield.

\section{Comparison of Monte Carlo simulations}
\label{sec:comparison}

A series of comparisons between our MC algorithm and Geant4 has been reported in previous
works~\cite{AstroPh,UHECR11,ICRC11}, showing general agreement. However, non-negligible discrepancies were found and
they have been being investigated. As a result of these studies, several improvements have been made in our MC
algorithm:

\begin{enumerate}

\item Corrections for the density effect in molecular cross sections, only noticeable at very high energy, have
    been reviewed and now are applied to the cross section for K-shell ionization as well (see~\cite{Thesis} for
    details).

\item The energy distribution of secondary electrons has been modified to better reproduce the end tail of $\delta$
    rays given by M\"{o}ller scattering.

\item Angular-differential cross sections (both elastic and inelastic) have been corrected for some relativistic
    effects.

\end{enumerate}

Updated results from our MC algorithm are presented here and compared with results from Geant4. In
figure~\ref{fig:cube}, a comparison for the average energy deposition per electron in a cube of 10~cm side filled with
air at 1.29~g/cm$^3$ (1013~hPa and 273~K) is shown. The geometry was defined such that electrons crossed the cube along
an axis going through the center of opposite faces. For this comparison, the standard physics list G4emStandard$\_$opt3
of Geant4 version 9.4, including multiple scattering, ionization, bremsstrahlung and emission of both X rays and Auger
electrons, was used. An excellent agreement ($<1\%$) holds between both simulations in the wide energy range from 1~MeV
to 100~GeV. Different geometries and other physics lists of Geant4 have been tested, always showing excellent
agreement.

In~\cite{ICRC11}, we reported some minor disagreement ($\sim2\%$) between our MC algorithm and Geant4. They are mainly
attributed to the inaccuracies in the energy distribution of secondaries in earlier versions of our code
(improvement~2). The corrections in the angular-differential cross sections (improvement~3) turned out to be
appreciable only at low energy ($\lesssim1$~MeV), since high-energy electrons follow almost rectilinear trajectories at
those conditions. The latter corrections also contribute to a better agreement between both simulations.

\begin{figure}
\centering
\includegraphics[width=0.32\linewidth]{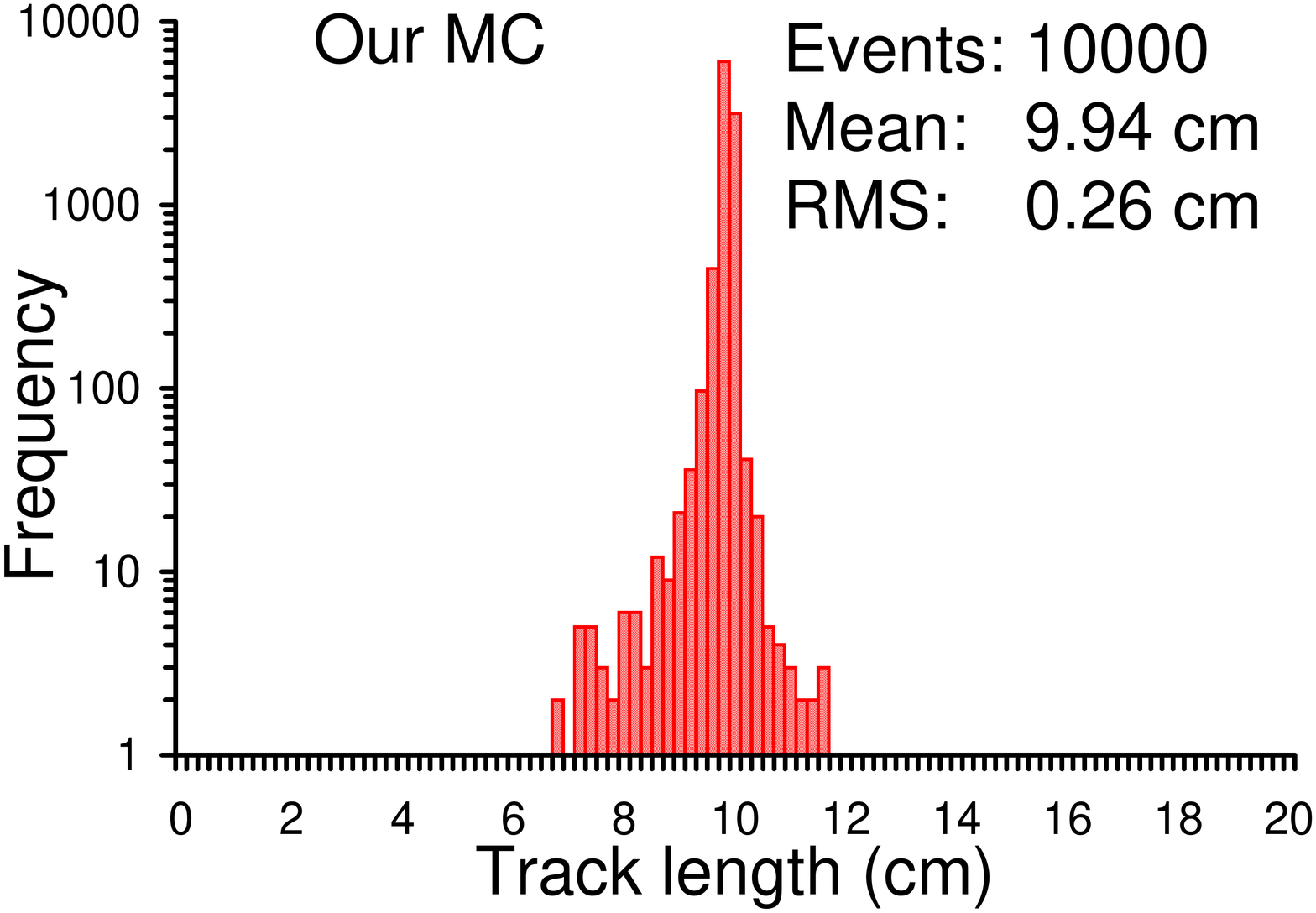}
\includegraphics[width=0.32\linewidth]{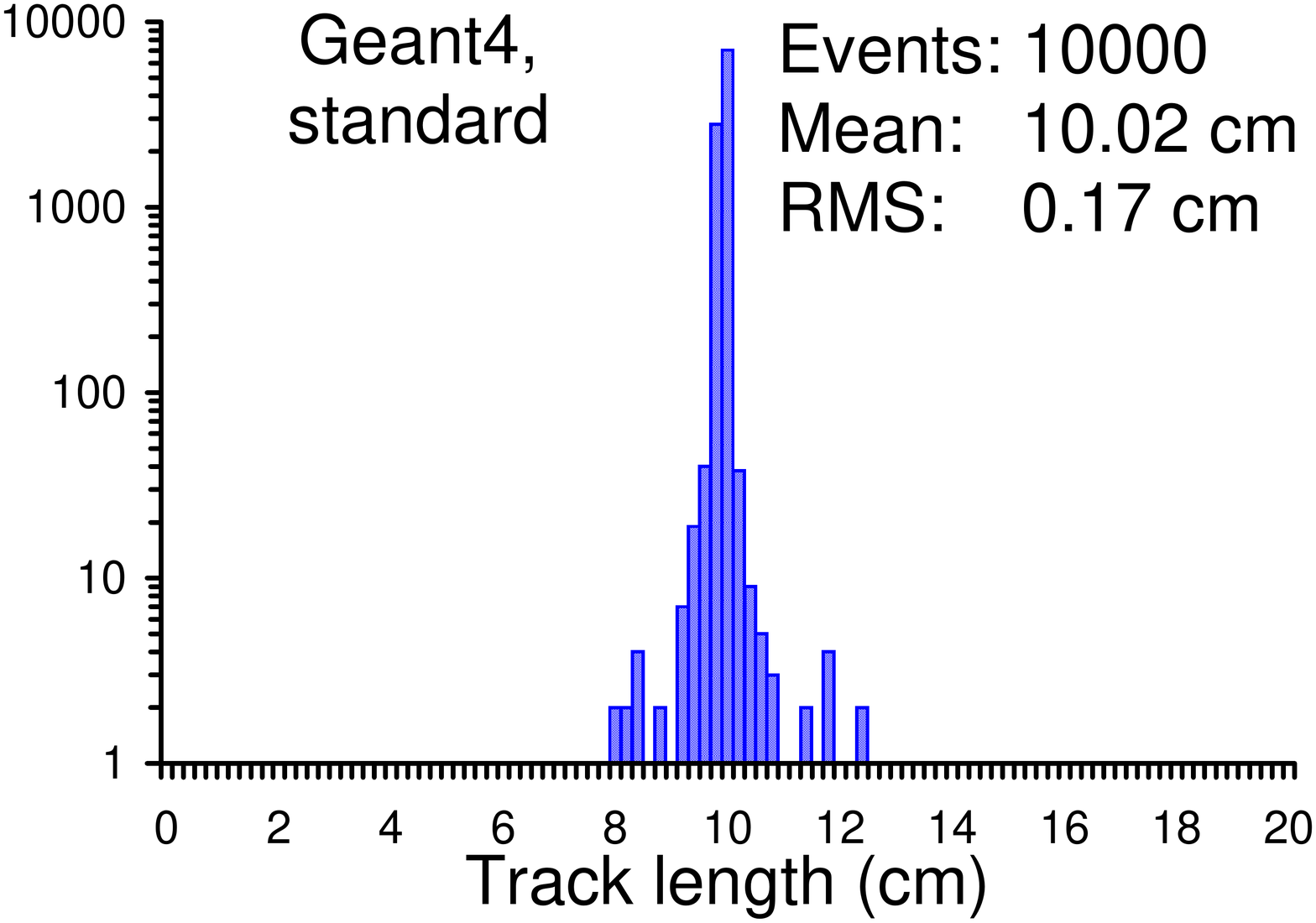}
\includegraphics[width=0.32\linewidth]{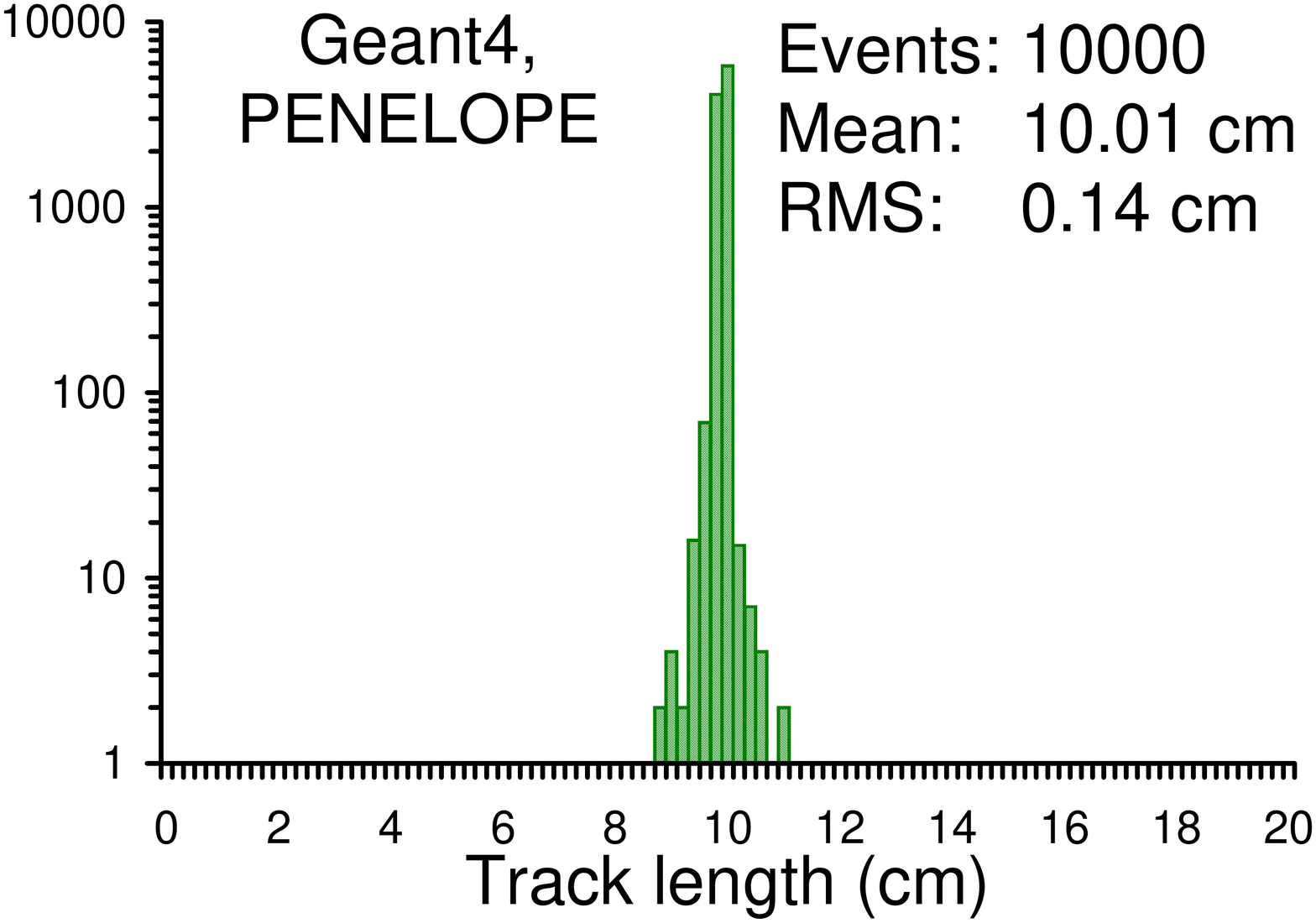}
\centering
\caption{%
Track lengths of 1~MeV electrons going through a 10~cm sphere (air at atmospheric pressure).
From left to right, histograms are obtained using our MC algorithm, Geant4 with the standard physics list and Geant4
with the low-energy PENELOPE extension.
} \label{fig:track_length}
\end{figure}

On the other hand, we still find some differences between our MC algorithm and Geant4 in the simulated trajectories of
electrons, which may be relevant at keV energies or in experiments with strong dependencies on geometrical factors. In
figure~\ref{fig:track_length}, histograms of track lengths of 1~MeV electrons crossing a 10~cm diameter sphere filled
with air at atmospheric pressure are shown for data from our upgraded MC algorithm and from Geant4 with the standard
physics list or the low-energy PENELOPE extension (G4emPenelope library). Notice that data from our MC algorithm are
more spread and its mean value is slightly lower than the predictions from both Geant4 simulations. These discrepancies
could be due to the approximation of multiple scattering employed by Geant4 and are under study.

\begin{figure}
\centering
\includegraphics[width=0.65\linewidth]{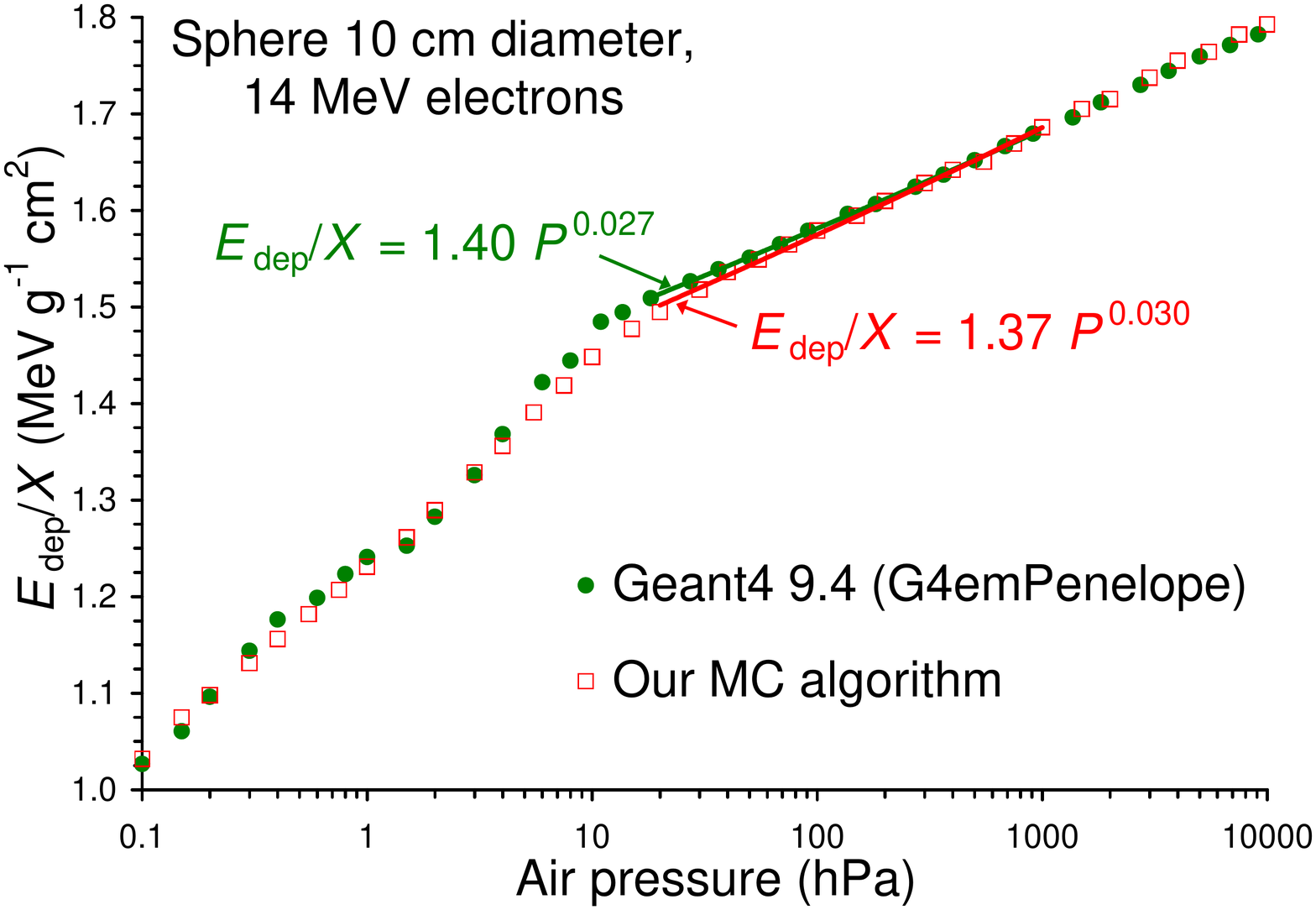}
\caption{Comparison between our MC algorithm and Geant4 for the pressure dependence of the energy deposition per unit
mass thickness of air for 14~MeV electrons crossing a 10~cm diameter sphere.
Both simulations predict a power law with a marked change of slope at around 10~hPa.
Best fits of data in the 15 $-$ 1000~hPa range are also shown.}
\label{fig:sphere}
\end{figure}

Comparisons between our MC algorithm and Geant4 for the pressure dependence of the energy deposition have also been
carried out. For this purpose, the energy deposited per unit mass thickness (MeV~g$^{-1}$~cm$^2$) is calculated as the
ratio of the average energy deposition and the average mass thickness $X$ traveled by incident electrons. Note that the
trivial linear growth of the total energy loss with mass thickness vanishes in this ratio and it solely remains the
pressure dependence due to the escaping energy carried by secondary radiation. This parameter is also insensitive to
possible slight differences between the simulated track lengths of incident electrons. In figure~\ref{fig:sphere},
results of the $E_{\rm dep}/X$ ratio for 14~MeV electrons crossing a 10~cm sphere are shown for the 0.1 $-$ 10000~hPa
air pressure range (293~K). Very good agreement between both simulations is observed again, obtaining an approximated
power law with a marked change of slope at around 10~hPa, which is associated to the average range of 410~eV photons
emitted by nitrogen after K-shell ionization (see~\cite{Thesis} for a thorough interpretation of simulation results).
In this comparison, the low-energy PENELOPE extension of Geant4 has been used instead, as it allows to reduce the cut
in energy for production of secondary radiation down to a value as low as 10~eV (the lower the pressure, the lower the
energy of secondary particles that can escape the sphere). The standard implementation of Geant4 gives very similar
results at pressures above 15~hPa. Best fits of data in the 15 $-$ 1000~hPa interval (typical relevant pressures for
cosmic-ray detection) are also shown in the figure.

It is worth mentioning that our Geant4 simulations are fully consistent with those performed by the
MACFLY~\cite{MACFLY} and AIRFLY~\cite{AIRFLY_P} experiments, which used this MC code to evaluate the energy deposition
in their collision chambers. Some differences with respect to Geant4 calculations made by the AirLight
experiment~\cite{AirLight} are found, nevertheless the authors of this experiment reported very
recently~\cite{AirLight_8thAFW} a bug in their simulation that likely justifies these discrepancies. Dedicated
comparisons with other MC codes have not been performed yet, however tests have also been made~\cite{AstroPh,UHECR11}
using EGS4 data given by the FLASH experiment for 28.5~GeV electrons~\cite{FLASH}. These calculations deviate more
significantly from ours (up to $8\%$, depending on pressure), although the discrepancies could fairly be explained by a
possible inaccurate treatment in~\cite{FLASH} of the density effect, which is very significant at such an energy.
Geometrical factors related to the spatial distribution of energy deposition in this experiment have also been
investigated, finding some discrepancies between the simulations that are still unclear.

Taking into consideration that our MC algorithm and Geant4 use different approaches for particle transport and are
based on nearly independent molecular data, this consistency of results suggests that a high-precision level in the
energy deposition is reached. On the other hand, this precision is limited by the accuracy in reference data of the
Bethe-Bloch stopping power, which is estimated to be about 1\% in the energy range of interest~\cite{ICRU37}. In view
of that, we estimate a conservative total uncertainty of 2\% in the above calculations of energy deposition.

\section{Impact on the fluorescence yield}
\label{sec:FY}

Several absolute measurements of the air-fluorescence yield are available in the literature. In recent
experiments~\cite{MACFLY,FLASH,AirLight,AIRFLY_8thAFW}, energy deposition was carefully evaluated by means of detailed
Geant4 or EGS4 simulations. However, the authors of other well-known experiments~\cite{Kakimoto,Nagano,Lefeuvre}
determined it from the Bethe-Bloch formula ignoring the effect of the escaping secondary radiation, which can cause
large systematic errors, as discussed above.

In previous works~\cite{AstroPh,UHECR11,ICRC11}, we proposed corrections to the energy deposition calculated in some of
these experiments and the fluorescence-yield values were re-normalized correspondingly (the experimental fluorescence
intensities were unchanged). As a consequence of these corrections, a very consistent sample of fluorescence-yield
values was obtained, allowing us to calculate a reliable world average of this important parameter. The above-mentioned
upgrades in our MC algorithm slightly affect this average. Next, updated results from this analysis are presented.

\begin{figure}
\centering
\includegraphics[width=0.49\linewidth]{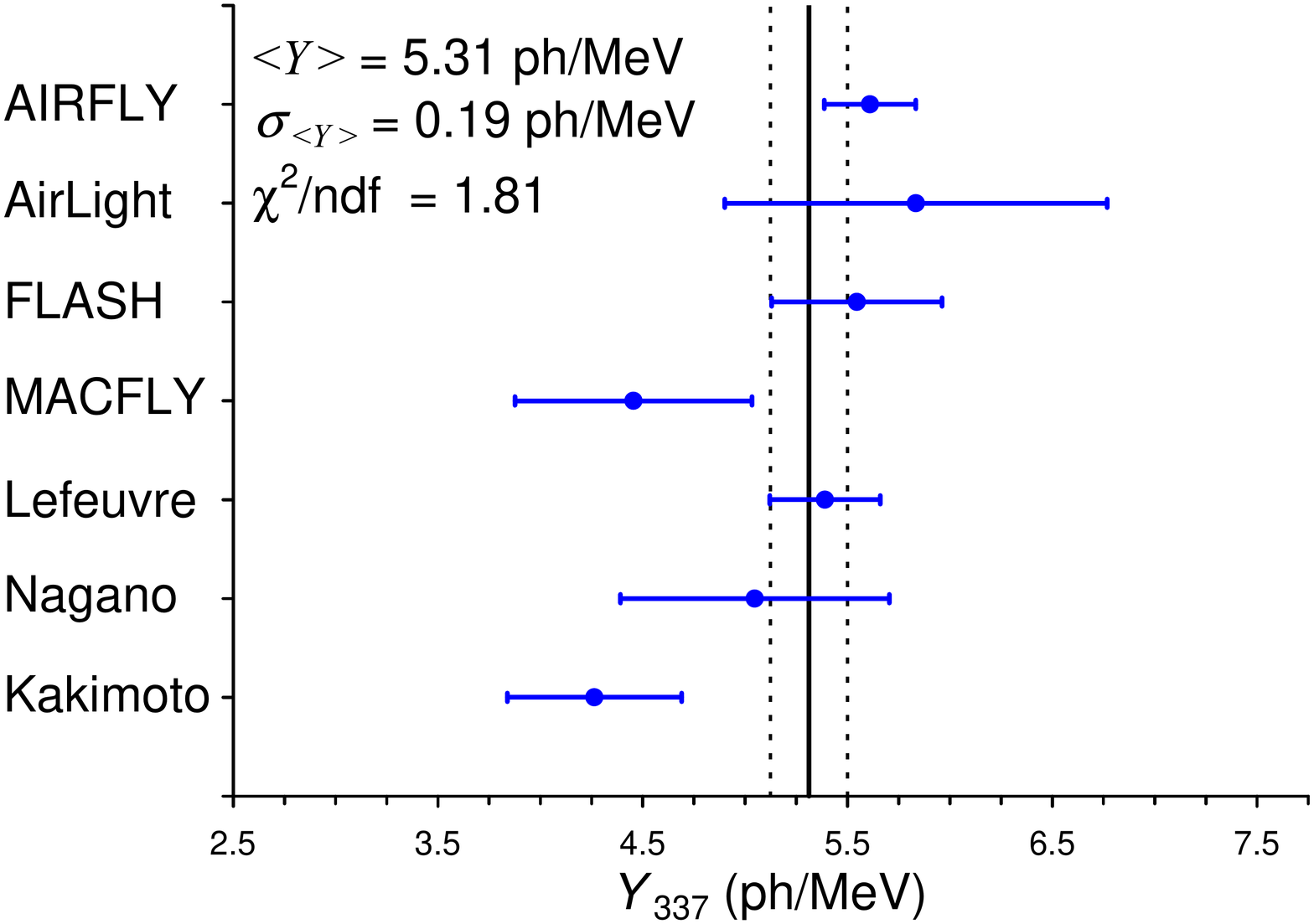}
\includegraphics[width=0.49\linewidth]{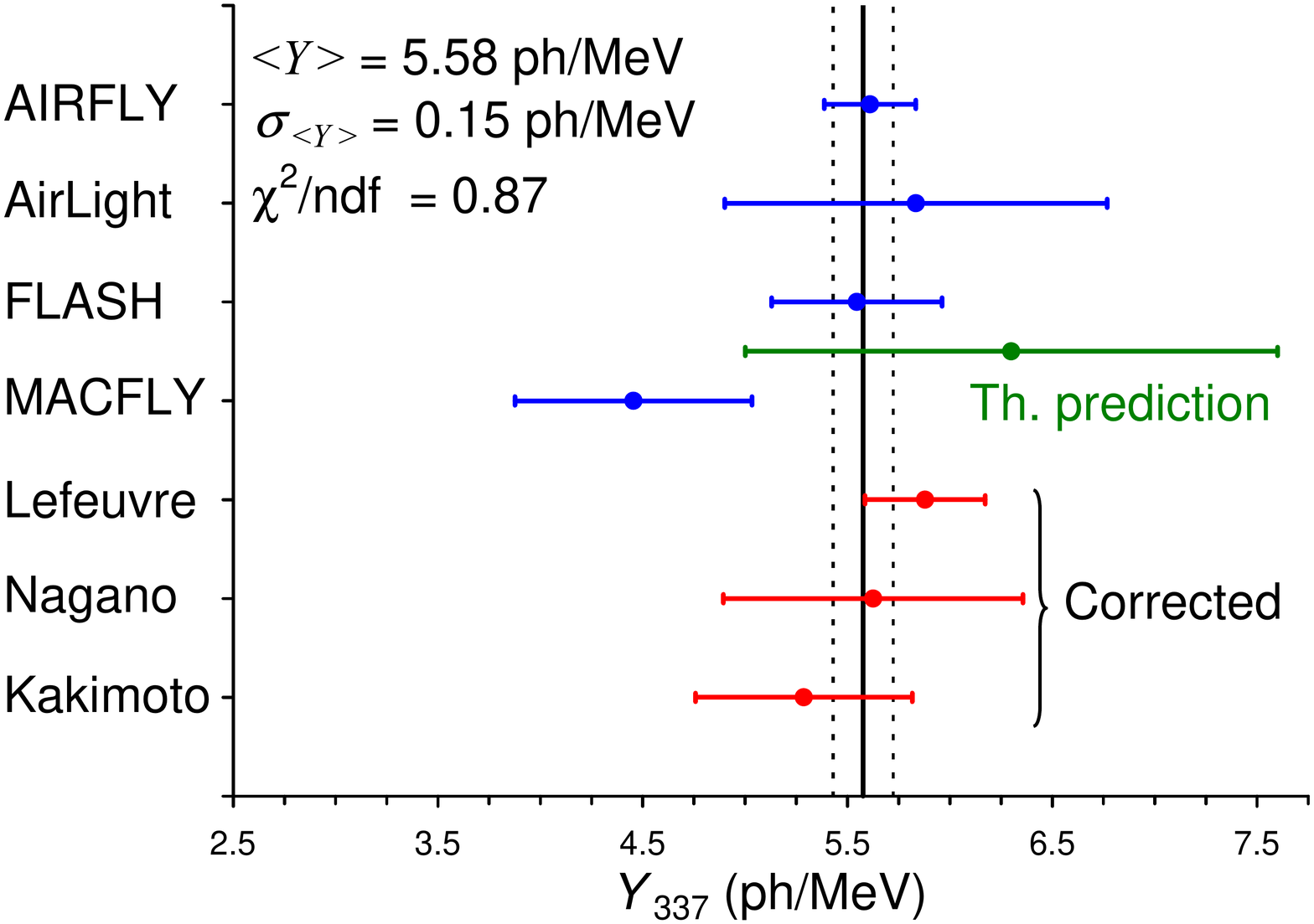}
\centering
\caption{%
Effect of our corrections on the available fluorescence-yield measurements normalized to common conditions (1013~hPa,
293~K and 337~nm band).
Original results are on the left and corrected ones (when applicable) are on the right, showing better compatibility.
Vertical lines represent the position of the sample weighted average $\langle Y\rangle$ and its standard error
$\sigma_{\langle Y\rangle}$.
The theoretical fluorescence-yield value predicted by our MC simulation is also shown for comparison.
} \label{fig:FY}
\end{figure}

Figure~\ref{fig:FY} shows a comparison of the available experimental data of the absolute fluorescence yield normalized
to common conditions, i.e., 337~nm band, dry air at 1013~hPa and 293~K.\footnote{Quenching data as well as intensities
of the remaining bands relative to the 337~nm band necessary for this normalization have been taken
from~\cite{AIRFLY_P}.} Original fluorescence yields (after being normalized to these reference conditions) are plotted
on the left. In the right plot, results from experiments that did not evaluate the energy deposition by simulation are
corrected according to our calculations. Experimental uncertainties are not modified in any case, and it has been
assumed that different measurements of a same experiment (only mean values are shown in~\ref{fig:FY}) are fully
correlated but no correlation exists between experiments. The sample includes the very precise measurement recently
reported by the AIRFLY Collaboration~\cite{AIRFLY_8thAFW}, that use Geant4 to evaluate the energy deposition assuming
an error contribution of 2\% to the total uncertainty of 4\%.

Our corrections increase the fluorescence yields of~\cite{Kakimoto,Nagano,Lefeuvre} by $9-24\%$, leading to an improved
compatibility of results. Notice that the $\chi^2$ statistic normalized by the number of degrees of freedom is lowered
from 1.81 to 0.87 after our corrections. The sample average $\langle Y\rangle=5.58$~ph/MeV is calculated using as
weights the reciprocals of the squared experimental uncertainties. Somewhat different averages ranging from 5.46~ph/MeV
to 5.68~ph/MeV are derived from plausible variations on the statistical analysis (e.g., treatment of outliers, weights
and correlations) and on our corrections (e.g., geometrical implementation details in our simulations). The standard
error of the weighted mean $\sigma_{\langle Y\rangle}=0.15$~ph/MeV (2.6\%) is determined from the quoted uncertainties
following the usual procedure. However, this low uncertainty seems to us to be unrealistic, because some experiments
did not include any error contribution from the evaluation of the energy deposition, which we estimate to be 2\% at
least for Geant4 and our MC algorithm. In addition, there are still some discrepancies between simulations to be
elucidated, as pointed out above. Taking into account all these considerations, we propose an average
fluorescence-yield value of $5.58$~ph/MeV and a conservative uncertainty estimate of 4\%, i.e., as low as the smallest
uncertainty of the sample corresponding to the AIRFLY measurement. Indeed, it can be understood from the present
analysis that this average and the AIRFLY measurement ($5.61$~ph/MeV) are complementary results showing high
reliability.

Our MC algorithm is able to simulate both the energy deposition and the fluorescence emission~\cite{NJP,Thesis}, which
in combination with experimental quenching data from~\cite{AIRFLY_P} also allows us to give a prediction of the
absolute fluorescence yield of $6.3\pm1.3$~ph/MeV, in agreement with experimental data (figure~\ref{fig:FY}, right
plot). Although we estimate a large uncertainty in this value due to the uncertainties in the many molecular parameters
involved in the calculation, the agreement with the experimental results provides a valuable theoretical support to
these measurements.

\section{Conclusions}
\label{sec:3}

We show that effects of secondary radiation escaping the collision chamber are very relevant for a correct
determination of the energy deposition in experiments dedicated to measure the air-fluorescence yield. Assuming the
Bethe-Bloch stopping power instead of computing the energy deposition by simulation may lead to large systematic errors
in the fluorescence yield.

New results of energy deposition from an upgraded MC algorithm that we have developed are presented here, showing an
excellent agreement with Geant4 for electrons with energies in the wide 1~MeV $-$ 100~GeV range. From the present
comparison, we conclude that these calculations have a precision level of 2\%. On the other hand, small differences in
the simulated electron trajectories are found, with possible implications at lower energy or in experiments with strong
dependencies on geometrical factors.

Our updated calculations of energy deposition have been used to correct experimental fluorescence-yield values reported
by experiments that did not evaluate the energy deposition by simulation. These corrections improve significantly the
compatibility of the available measurements of the absolute air-fluorescence yield. As a result of this analysis, we
present a reliable average of 5.58~ph/MeV with an estimated uncertainty of 4\%. In addition, experimental data are
shown to be consistent with a theoretical prediction of the absolute fluorescence yield that we obtained by means of
our MC simulation.

\section*{Acknowledgements}
This work has been supported by the Spanish Ministerio de Ciencia e Innovaci\'{o}n (FPA2009-07772 and CONSOLIDER CPAN
CSD2007-42) and ``Comunidad de Madrid'' (Ref.: 910600).

\end{document}